%% file: root.tex
\documentclass[10pt, conference, compsocconf]{IEEEtran}
\ifCLASSINFOpdf
\else
\fi
\usepackage{caption}
\usepackage{graphicx}
\usepackage{epstopdf}
\usepackage{url}
\usepackage{proof}
\usepackage{framed}
\usepackage{multicol}
\usepackage{etaremune}
\usepackage{listings}
\usepackage{hyperref}
\usepackage[inference]{semantic}
\usepackage{amsmath}
\usepackage{amsthm,amssymb,amsfonts}
\usepackage{paralist}
\usepackage{color}
\usepackage{pdflscape}
\usepackage{xspace}
\usepackage{bookmark}
\usepackage{algorithm}
\usepackage{algpseudocode}
\usepackage{subcaption}
\usepackage{enumitem}
\usepackage{todonotes}
\usepackage{cleveref}
\usepackage{tikz}
\usetikzlibrary{arrows,positioning,tikzmark}
\usepackage{flushend}

\newcommand{\ignoreA}[1]{}

\newcommand{\N}{\mathbb{N}}

\newcommand{\specialcell}[2][c]{%
  \begin{tabular}[#1]{@{}c@{}}#2\end{tabular}}

\DeclareMathSizes{14}{14}{14}{14}

\begin{document}
%
\title{An Operational Semantic Basis\\for OpenMP Race Analysis}


\author{\IEEEauthorblockN{Simone Atzeni}
\IEEEauthorblockA{School of Computing\\
University of Utah\\
Salt Lake City, UT, USA\\
simone@cs.utah.edu}
\and
\IEEEauthorblockN{Ganesh Gopalakrishnan}
\IEEEauthorblockA{School of Computing\\
University of Utah\\
Salt Lake City, UT, USA\\
ganesh@cs.utah.edu}
}


%


\input{macros.tex}

\maketitle

\input{abstract.tex}

\begin{IEEEkeywords}
  OpenMP; operational semantics; concurrency; formal definition; data race;
  data race detection tool; structured parallelism
\end{IEEEkeywords}

%
\IEEEpeerreviewmaketitle

\input{introduction}
\input{background}
\input{operationalsemantics}
\input{implementation}
\input{conclusions}

\bibliographystyle{IEEEtran}
\bibliography{references}



%



\end{document}

%% file: macros.tex
\newcommand*{\DEBUG}{}
\ifdefined\DEBUG
\newcommand{\simone}[1]{\todo[inline,size=\small, color=green!40]{S: #1}}
\else
\newcommand{\simone}[1]{}
\fi
\newcommand{\ganesh}[1]{\todo[inline,size=\small, color=yellow!40]{G: #1}}

\newcommand{\comment}[1]{\todo[inline,size=\tiny, color=blue!40]{comment out: #1}}

\newcommand{\zvonimir}[1]{\todo[inline,size=\small, color=blue!40]{Z: #1}}
\newcommand{\dong}[1]{\todo[inline,size=\small, color=red!40]{D: #1}}
\newcommand{\ignacio}[1]{\todo[inline,size=\small, color=yellow!40]{I: #1}}
\newcommand{\greg}[1]{\todo[inline,size=\small, color=yellow!40]{L: #1}}
\newcommand{\martin}[1]{\todo[inline,size=\small, color=yellow!40]{M: #1}}

\newcommand{\archer}{\textsc{Archer}\xspace}
\newcommand{\sword}{\textsc{Sword}\xspace}
\newcommand{\tsan}{TSan\xspace}
\newcommand{\iomp}{Intel\circledR OpenMP* Runtime\xspace}
\newcommand{\omprt}{LLVM OpenMP Runtime\xspace}
\newcommand{\insp}{Intel\circledR Inspector XE\xspace}
\newcommand{\amg}{AMG2013\xspace}
\newcommand{\omp}{OpenMP\xspace}


%% file: abstract.tex
\begin{abstract}
  OpenMP is the de facto standard to exploit the on-node parallelism in new
  generation supercomputers.
  Despite its overall ease of use, even expert users are known to create
  OpenMP programs that harbor concurrency errors, of which one of the most
  insidious of errors are {\em data races}.
  OpenMP is also a rapidly evolving standard, which means that future data
  races may be introduced within unfamiliar contexts.
  A simple and rigorous operational semantics for OpenMP can help build
  reliable race checkers and ward off future errors through programmer
  education and better tooling.
  This paper's key contribution is a simple operational semantics for
  OpenMP, with primitive events matching those generated by today's
  popular OpenMP runtimes and tracing methods such as OMPT.
  This makes our operational semantics more than a theoretical document for
  intellectual edification; it can serve as a blueprint for OpenMP event
  capture and tool building.
  We back this statement by summarizing the workings of a new data race
  checker for OpenMP being built based on this semantics.
  The larger purpose served by our semantics is to serve the needs of the
  OpenMP community with regard to their contemplated extensions to OpenMP, as
  well as future tooling efforts.
\end{abstract}

%% file: introduction.tex
\section{Introduction}
\label{sec:introduction}

OpenMP is the de facto standard for on-node parallelism in High Performance
Computing.
While OpenMP is highly portable and easy to use, it is also error-prone.
Data races are one of the major source of errors in OpenMP based HPC
applications.
Although many existing correctness checking tools support programmers in the
detection and removal of data races, these tools rely on static and dynamic
analyses that either may not fit well with the needs of practical OpenMP race
checking~\cite{Savage:1997:EDD:269005.266641,flanagan_fasttrack:_2009,Chatarasi2017},
miss races, or incur high overheads.
Even symbolic analysis methods for OpenMP suffer from these
issues~\cite{DBLP:conf/icpp/MaDWLQY13}.

Our past work has successfully adapted static and dynamic analysis to OpenMP
and offered a practical race checker called \archer that has caught data races
in critical field applications~\cite{atzeni_archer:_2016}.
However, \archer suffers from high memory overheads, and misses races in many
cases due to its exclusive reliance on the {\em happens before} model.
It is well known that the races caught under this model depend on the schedule
actually played out.
That is, races within alternate schedules may be missed.

Our approach is to follow the lead of those who have exploited structured
parallelism to make race-checking simpler and more efficient--e.g., for
OpenMP~\cite{tim_lewis_openmp_nodate}, Cilk~\cite{cilk}, and
X10~\cite{Charles:2005:XOA:1103845.1094852}.
We define an operational semantics that models the concurrency structure of
OpenMP programs, exploiting the tool API (OMPT)~\cite{eichenberger_ompt:_2013}
of modern OpenMP runtimes to identify every OpenMP event in the execution.
Our approach also has the flavor of combining the exploitation of structured
parallelism with lock-set based race checking (see
Section~\ref{sec:ruleswalkthrough} for our lock handling rules).
The result is a more precise and traceable data race checker based on a clear
operational semantics that fits in one page over 10 rules
(Section~\ref{sec:ruleswalkthrough}), supported by some helper functions
(Section~\ref{sec:helperspredicates}).
We believe that our formalization will benefit designers who seek to model new
data race detection techniques for structured parallelism (in particular
OpenMP) and those seeking to build and understand new and existing data race
checkers.

\input{interleavings}

While Raman et.\ al.'s work in this
area~\cite{Raman:2012:SPD:2345156.2254127,Raman:2010:EDR:1939399.1939430}
proposes techniques to exploit the structured parallelism on parallel
programming models such as Cilk and X10, their techniques currently are
focused on async/finish structured parallelism of X10 and Habanero-Java.
This makes their technique not directly applicable to OpenMP at this point.
%
%
\noindent To summarize, the main contributions of this work are:

\begin{itemize}
\item An operational semantics that model the concurrency structure of an
  OpenMP program matching the OMPT events, and an overview of a prototype race
  checker that demonstrates how such a semantics can be a workhorse for race
  checking.
\item A set of rules that exploit the OpenMP structured parallelism to
  identify races.
\item An extensible operational semantics that allows future OpenMP constructs
  to be captured and analyzed.
\end{itemize}

The remainder of this paper is structured as follows:
Section~\ref{sec:background} discusses limitations of existing techniques that
our operational semantics can overcome; Section~\ref{sec:operationalsemantics}
illustrates the state machine that implements the operational semantics rules,
the conventions used to define the operational semantics rules, how we model
the OpenMP constructs in our concurrency model, and a real example to show the
effectiveness of the operational semantics in identifying data races;
Section~\ref{sec:implementation} gives some ideas of a possible implementation
of the operational semantics in a real data race detection tool;
Section~\ref{sec:conclusions} concludes the paper.


%% file: interleavings.tex
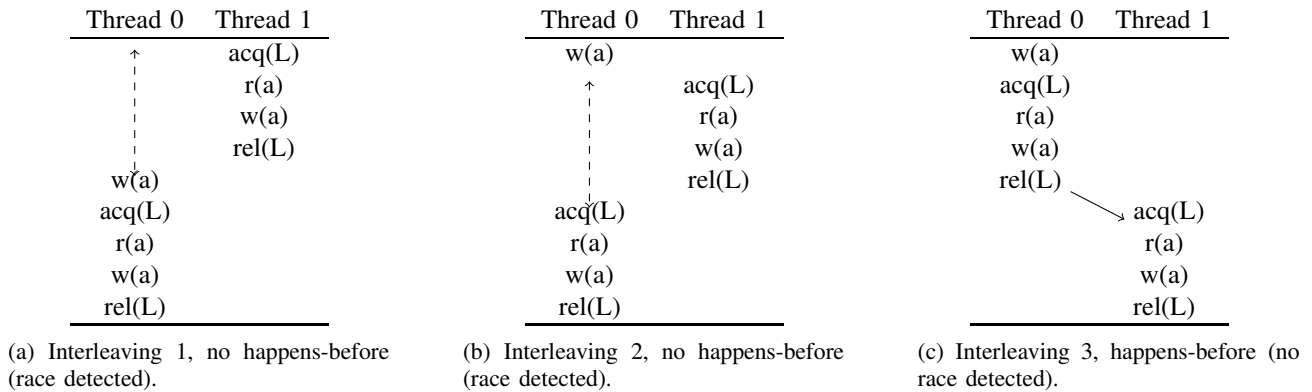
\begin{figure*}[t]
  \hfill
  \begin{subtable}{.28\textwidth}
    \centering
    \begin{tabular}{c c}
      Thread~0 & Thread~1 \\
      \hline
      \tikzmark{a} & acq(L) \\
               & r(a) \\
               & w(a) \\
      \tikzmark{b} & rel(L) \\
      w(a)   & \\
      acq(L) & \\
      r(a)   & \\
      w(a)   & \\
      rel(L) & \\
      \hline
    \end{tabular}
    \subcaption{Interleaving 1, no happens-before (race detected).}
    \label{fig:interleaving1}
  \end{subtable}
  \hfill
  \begin{subtable}{.28\textwidth}
    \centering
    \begin{tabular}{c c}
      Thread~0 & Thread~1 \\
      \hline
      w(a)     & \\
      \tikzmark{c} & acq(L) \\
               & r(a) \\
               & w(a) \\
      \tikzmark{d} & rel(L) \\
      acq(L) & \\
      r(a)   & \\
      w(a)   & \\
      rel(L) & \\
      \hline
    \end{tabular}
    \subcaption{Interleaving 2, no happens-before (race detected).}
    \label{fig:interleaving2}
  \end{subtable}
  \hfill
  \begin{subtable}{.28\textwidth}
    \centering
    \begin{tabular}{c c}
      Thread~0 & Thread~1 \\
      \hline
      w(a)   & \\
      acq(L) & \\
      r(a)   & \\
      w(a)   & \\
      rel(L) \tikzmark{e} & \\
               & \tikzmark{f} acq(L) \\
               & r(a) \\
               & w(a) \\
               & rel(L) \\
      \hline
    \end{tabular}
    \begin{tikzpicture}[overlay, remember picture, yshift=.25\baselineskip, shorten >=-7.5pt, shorten <=-3.5pt]
      \draw[dashed,<->] ([yshift=.75pt]{pic cs:a}) -- ({pic cs:b});
      \draw[dashed,<->] ([yshift=.75pt]{pic cs:c}) -- ({pic cs:d});
    \end{tikzpicture}
    \begin{tikzpicture}[overlay, remember picture, yshift=0.25\baselineskip, shorten >=0pt, shorten <=0pt]
      \draw[->] ([yshift=-1.5pt]{pic cs:e}) -- ({pic cs:f});
    \end{tikzpicture}
    \subcaption{Interleaving 3, happens-before (no race detected).}
    \label{fig:interleaving3}
  \end{subtable}
  \hfill
  \caption{Possible interleavings for program in
    Listings~\ref{code:example01}. The dashed line indicates the write
    operation of thread 0 can happen simultaneously with the operations of
    thread 1. The solid line indicates the happens-before edge between the
    threads.}
  \label{fig:interleavings}
  \vspace{-5pt}
\end{figure*}


%% file: background.tex
\section{Background}
\label{sec:background}

In this section we give an overview of the {\em happens-before relation} for
dynamic data race detection analysis that underlies existing race
detectors~\cite{atzeni_archer:_2016,flanagan_fasttrack:_2009,OCallahan:2003:HDD:966049.781528}.
For our purposes, event $a$ happens
before~\cite{time-clocks-ordering-events-distributed-system} event $b$
($a \rightarrow b$) if (1)~they occur in that order within the same thread,
(2)~if $a$ is an unlock and $b$ is a lock, or (3)~they are synchronized
otherwise (e.g., $a$ is before a barrier and $b$ happens after the barrier).
A data race is a happens-before unordered pair of events where one event is a
write.
Vector-clocks~\cite{Mattern88virtualtime,fidge_vector_clocks_1988} and their
adaptations~\cite{flanagan_fasttrack:_2009} typically help realize
happens-before.
Happens-before is defined {\em per thread schedule}, thus making
happens-before based race detectors miss races when they do not exercise all
schedules.
For example, in listing~\ref{code:example01}, we depict a parallel region with
two threads.
The main thread initializes $a$ inside the master construct, and both threads
write variable $a$ within a critical section.
Because the OpenMP master construct does not enforce an implicit barrier at
its termination point, while thread $0$ initializes $a$, the thread $1$ can
simultaneously access $a$ within the critical section, introducing a data
race.

In Figure~\ref{fig:interleavings} we exhibit three different thread
interleavings for the program in Listing~\ref{code:example01}.
In the first two interleavings, the data race on \emph{a} manifests itself.
Indeed in Figure~\ref{fig:interleaving1}, first thread $2$ reads and writes
within a synchronization block, while thread $1$ performs a non-synchronized
write.
As shown in the figure, the non-synchronized write from thread $1$ can happen
anytime, even though thread $2$ is accessing $a$ within a critical section.
In Figure~\ref{fig:interleaving2}, first thread $1$ performs a
non-synchronized write and thread $2$ reads and writes within a
synchronization block.
In both cases, the two threads access simultaneously \emph{a} and the data
race detection algorithm shows the absence of happens-before between the
threads, catching the data race.
On the other hand, in Figure~\ref{fig:interleaving3} we have the typical
situation where happens-before masks a race.
Thread $1$ executes both non-synchronized and synchronized accesses on $a$
before thread $2$ performs any other operation.
The release of the lock by thread $1$ creates a happens-before edge with the
acquiring of the same lock by thread $2$, masking the previous
non-synchronized write by the first thread.

In our approach, races such as in Figure~\ref{fig:interleavings} are detected
thanks to a global data structure that maintains relevant memory accesses
information performed by the threads, along with other information such as
operation type, thread id, and locks held while making accesses.
At each barrier, the operational semantics verifies the presence of data
races, analyzing all the memory accesses performed by the threads up to that
point, ensuring no data race will be missed (details are in
Section~\ref{sec:operationalsemantics}).

\input{example01}

\begin{figure*}[t]
  \centering
  \includegraphics[width=0.7\textwidth]{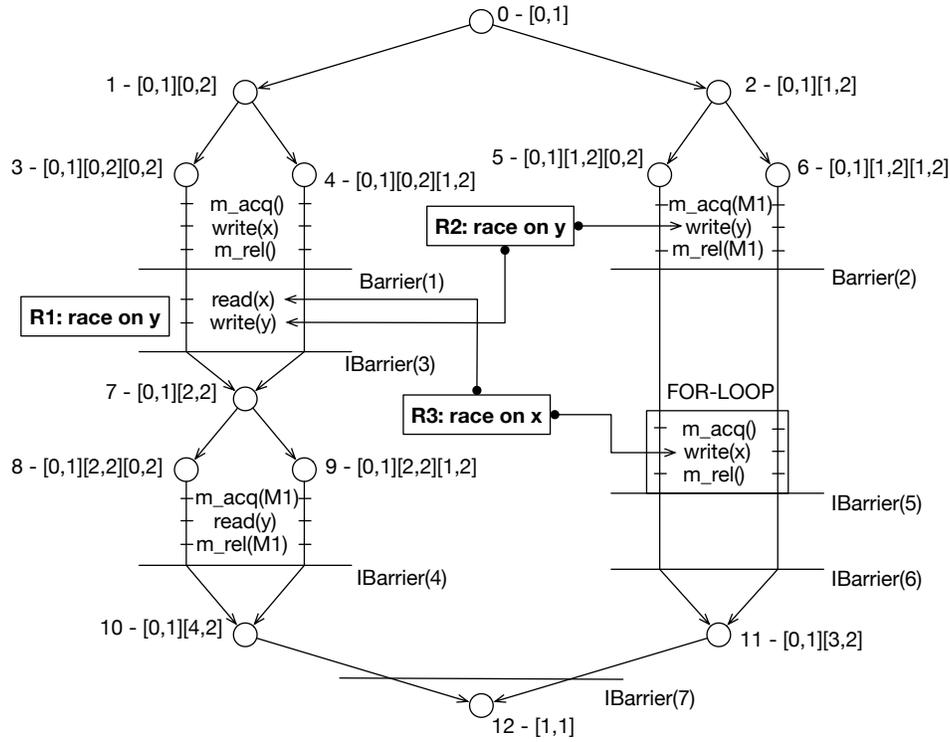}
  \caption{Structure of the OpenMP program in Listing~\ref{code:example04}.}
  \label{fig:example04}
\end{figure*}

A key property of our operational semantics is that it highlights the
concurrency structure created by a particular OpenMP program.
If a particular thread forks two different threads and these threads perform
their own accesses, our semantics records these accesses {\em not in terms of
  a particular interleaving}, but as a pair of accesses at {\em specific
  positions in the fork-join structure}, together with the {\em mutex locks
  held} when making the access.
We exploit the idea of {\em offset span labels} pioneered by
Mellor-Crummey~\cite{Mellor-Crummey:1991:ODD:125826.125861} to record
``positions'' within the concurrency structure.
We believe that these mechanisms serve the dual purpose of (1)~creating a
concurrency representation that is general enough to ``hang'' on it future
extensions to OpenMP's concurrency structure, and (2)~also efficient enough to
support the creation of a dynamic race detector.


%% file: example01.tex
\lstset{language=C++,
  escapeinside=||,
  basicstyle=\ttfamily,
  showstringspaces=false,
  keywordstyle=\color{blue}\ttfamily,
  stringstyle=\color{red}\ttfamily,
  otherkeywords={omp, parallel, barrier, critical, num_threads, shared, master},
  breaklines=true,
  basicstyle=\small
}
\begin{lstlisting}[language=C++, caption=Data race in OpenMP program
that may not manifest at runtime., label=code:example01]
  int a;

#pragma omp parallel shared(a) num_threads(2)
  {
#pragma omp master
    {
      a = 0;
    }
#pragma omp critical
    {
      a += 1;
    }
  }
\end{lstlisting}


%% file: operationalsemantics.tex
\section{Operational Semantics}
\label{sec:operationalsemantics}

\begin{figure}[!t]
  \vspace{-9pt}
  \input{example04}
\end{figure}

The basic idea behind the operational semantics is to advance a state machine
along the execution of the program in response to OpenMP events, and update
the concurrency structure held in our state representation.
Typical events include fork/join events (begin/end of a parallel region),
acquiring and releasing of locks that guard critical sections, loads, stores,
etc.
The capturing of the OpenMP events is enabled by the new OpenMP Tools API
(OMPT)~\cite{eichenberger_ompt:_2013} that modern OpenMP runtime implements to
facilitate the development of correctness and performance tools.
The OMPT interface triggers a callback for each OpenMP event that happens at
runtime so that tools can access important information including parallel
regions creation, threads entering or exiting a critical section, barrier
executions, etc.
The operational semantics rules match the OMPT events to correctly represent
the concurrency structure of the OpenMP program.
Each thread maintains a label in terms of offset-span labels that marks its
{\em lineage} in the concurrency structure defined by prior forks and joins.
Figure~\ref{fig:example04} illustrates the concurrency structure of the code
in Listing~\ref{code:example04}, where circles represent the starting point of
threads, and vertical lines represent traces of a thread's execution.
Two or more diagonal lines that exit/enter the circles represent fork/join
points in the program.

In our example, master thread 0 creates a parallel region of two threads
(thread 1 and 2).
Each thread creates a nested parallel region of a team of two.
Because of the SIMD model followed,
{\em all} threads in a parallel region execute the operations indicated
at the horizontal tick marks.

Notice how each thread in the diagram has associated an id, and a label which
consists of pairs in square brackets.
The id identifies the thread in the diagram, while the second label
is the {\em offset-span label}.
The offset-span label length grows at each fork and
shrinks at each join.

When a thread reaches a fork, it creates a parallel region and a new pair of
integers is added to the offset-span label.
The first integer indicates the thread {\em rank} (ID) and the second one
indicates the number of the threads in the team.
On the other hand, when threads join, the last pair of the label is removed
and the previous label position is updated.
We do not provide all the details of offset-span label manipulations here (see
\cite{Mellor-Crummey:1991:ODD:125826.125861} for that); however, our semantic
rules do include all the relevant details (Section~\ref{sec:osl}).
Mellor-Crummey has shown that given two threads and their offset-span labels,
it is possible to determine if the two thread accesses are concurrent, and
this happens to be the crux of race checking.

In our example of Figure~\ref{fig:interleavings}, thread 0 creates the first
parallel region and the operational semantics records this event through one
of its rules.
The same happens for thread 1 and 2 when they create the two nested parallel
regions.
At this point, each thread starts the execution of the operations in the
program.
In both nested parallel regions, the threads acquire {\em different locks} to
access the shared variables.
This triggers specific operational semantic rules to record the operations in
the history of each thread.

More specifically, in the left parallel region, threads 3 and 4 enter a global
critical section, write on $x$ and exit from the critical section.
At the same time, threads 5 and 6 in the nested parallel region on the right
acquire a lock on $M1$, write on $y$ and release the lock.
Also, the loads and stores performed by threads trigger a rule that stores the
information about the memory accesses in a global structure along with the
thread id and the id of the mutexes previously acquired by the thread (if
any).

In our example, threads 3 and 4 reach the barrier 1 eventually, while threads
5 and 6 reach barrier 2.
When a parallel thread reaches a barrier (either implicit or explicit), it
waits for all the other threads in the team; they then synchronize and proceed
with the execution.
The state machine triggers different rules at the barrier to model the thread
synchronization--but more importantly {\em to perform the data race detection
  on the operations executed up to that point}.

The data race detection rule first identifies all possible concurrent threads
in the system, comparing their offset-span labels.
Second, it compares, for a given thread, its memory accesses with the memory
accesses of another concurrent thread.
If the rule identifies two memory accesses to a common location, at least one
write, and without synchronization (or different mutex ids), it reports the
race.\footnote{While these comparisons can make race-checking inefficient, our
  implementation in progress splits the burden into online event logging and
  offline event analysis that employs parallelism, as elaborated in
Section~\ref{sec:implementation}.}

Let us suppose the threads 8 and 9 have reached the implicit barrier 4, while
the threads 5 and 6 are waiting at the implicit barrier 6.
(Notice how threads 3 and 4 already joined into thread 7 which generated a new
nested parallel region with threads 8 and 9. The global data structure still
contains all the operations performed during the program execution up to those
barriers.)
All of the threads trigger the data race detection algorithm through one of
the barrier rules.
Up to that point, the global structure that collects the memory accesses
contains all the loads and stores executed by the threads and related mutex
information used for the memory accesses.
The data race algorithm has all the information to identify potential data
races.
As stated previously, the algorithm identifies and compares only the memory
accesses of concurrent threads.

In our example, there are three data races, identified by $R1$, $R2$, and
$R3$.
\begin{itemize}
\item $R1$ happens within the same nested parallel region on shared variable
  $y$.
  This happens because both thread 3 and 4 (that are concurrent) write the
  shared location without any synchronization.

\item Race $R2$ manifests between the threads of the two nested different
  parallel regions.
  The involved threads are 3 and 4 from the parallel region on the left, and 5
  and 6 from the right parallel region.
  All the threads are concurrent to each other: threads 5 and 6 write on $y$
  through the critical section $M1$ and they do not race with each other.
  However, the concurrent threads 3 and 4 write on the same shared variable
  without any synchronization racing with threads 5 and 6.

\item $R3$ is similar to thread $R2$ but on the shared data $x$.

\item The data race detection algorithm identifies the races by comparing all
  the memory accesses in the global structure only for the possible concurrent
  threads.
  It is interesting to notice that the algorithm does not report any races on
  $y$ between threads 3,4 and the threads 8,9.
  By comparing the offset-span labels, the algorithm recognizes that threads 3
  and 4 have already terminated when threads 8 and 9 start their work, so they
  are not deemed concurrent.
\end{itemize}

\noindent We now detail our semantics, presenting each of its component
building blocks in separate sections, followed by our semantic rules
themselves.

\subsection{Predicates and Conventions}
\label{sec:conventions}

We first need to state our conventions.
$\N$ is the set of natural numbers, $\{0,1,2,\ldots\}$.
$x\in\N$ can be treated as a set $\{0,\ldots,x-1\}$ as in set theory.
Thus, $0=\{\}$, $1=\{0\}$, $2=\{0,1\}$, $3=\{0,1,2\}$, etc.
Whenever we treat a member of $\N$ as a number as well as a set, we'll make
sure to provide a hint.
$t\in TID$ is a thread identifier for some $TID\in \N$.
$ADDR\in\N$ is the range of memory addresses accessed by the threads.

\input{rules}

\subsection{Offset-Span Labels}
\label{sec:osl}

We showed how the {\em offset-span labels} are used to identify whether two
threads are concurrent, and apply the data race detection only in that case.
The offset-span label mechanism was introduced
in~\cite{Mellor-Crummey:1991:ODD:125826.125861}.
An offset-span label, $osl$ for short, labels each thread's execution point
with a sequence of pairs, marking its lineage in the concurrency structure
defined by prior forks and joins.
The domain for the offset-span labels is $OSL = (\N\times\N)^{\N}$, i.e.\
each member $osl\in OSL$ is a sequence of pairs:

\[ [a_1,b_1][a_2,b_2],\ldots,[a_n,b_n].\]

Let us take two offset-span labels $osl_1, osl_2\in OSL$, respectively
associated to thread 1 and thread 2.
These labels are sequential (hence the thread 1 and thread 2 are not
concurrent) when:
\hfill \\
\paragraph*{\bf case 1} $\exists_{P,S}(osl_1=P) \wedge osl_2=PS$, where $P$ and $S$
are any non-null sequence of ordered label pairs.
\hfill \\
\paragraph*{\bf case 2}
$\exists_{P,S_x,S_y,o_x,o_y,s}(osl_1=P[o_x,s]S_x) \wedge osl_2=P[o_y,s]S_y)
\wedge (o_x < o_y) \wedge (o_x \mod s = o_y \mod s)$ where $P$, $S_x$, and $S_y$
are (possibly null) sequence of ordered pairs.
\hfill \\
\hfill \\
Otherwise, they are concurrent.

%
%
%
%

The offset-span label is an important piece of our concurrency model since it
gives precious information regarding whether two given threads can actually
race or not.
For further details, please see~\cite{Mellor-Crummey:1991:ODD:125826.125861}.

\subsection{System State}
\label{sec:systemstate}

The state of the system consists of a global state $GS$ and a set of thread
local states $TP$ (Thread Pool).
The total state $ts$ of any system is a pair ``Global State, Thread Pool''.
A specific total state $ts$ is:
\[ ts = \langle gs, tp \rangle \]
\noindent Each total state $ts$ originates from the domain $TS$, where
$TS = GS\times TP$.

\noindent Each global state $gs$ is a 4-tuple:
\[ \langle bm, m, rw, \sigma \rangle \]
\noindent Each global state $gs$ originates from the domain $GS$, where
\[ GS = BM\times M\times RW\times \Sigma \]

\noindent where:
\begin{itemize}
\item The domain $BM = ParRegID \mapsto (\N\times \N)$.
  Thus, for each $bm\in BM$, we have
  $bm\; :\; ParRegID \mapsto (\N\times \N)$.
  Given a $p\in ParRegID$, $bm$ returns a pair of natural numbers $(a,b)$,
  where $a$ is the ``current Barrier Count'' and $b$ is the ``target Barrier
  Count.''
  When a thread $t$ with offset-span label $osl$ executes a $ParBegin(N)$
  instruction, $N$ threads are created, and an entry
  $\langle osl, (0,N)\rangle$ is added to function $bm$\footnotemark.
  \footnotetext{Recall that functions are single-valued relations, or sets of
    pairs with unique second component for each given first component.
    Thus, $\{\langle osl,(0,N)\rangle\}$ is a function. We allow functions to
    evolve, i.e.\ undefined for items explicitly added.}
  The first field $a$ is incremented each time a thread hits a barrier.
  When the value reaches the number of threads in the team, it signals that
  all threads have synchronized at the barrier and the program can continue
  its execution.

\item ``Mutex'' $m$ comes from domain $M$ where
  $M = Names \mapsto (\{-1\} \cup TID)$.
  That is, given a mutex name $m \in Names$, $M[m] = -1$ means that this mutex
  is free.
  Otherwise, $M[m] = t$, recording the fact that this mutex is held by the
  task associated to thread $t$.
  We use the value $\mu$ to indicate a mutex that has no name associated.
  A mutex with no name is usually the common case in a OpenMP progam and it
  refers to any global critical section or lock (e.g.\ \texttt{\#pragma omp
    critical}).

\item Let memory access-type $MAT=\{R,W\}$ indicates a read or a write
  operation of a memory access.

\item $rw\in RW$ is a tuple (data structure) that maintains all the memory
  accesses of each thread in the system.
  We have
  $RW = TID\times OSL \times \N \times ADDR \times MAT \times M$.
  Each memory access performed by thread $t$ is recorded as the tuple
  \[ \langle tid, osl, bl, addr, mat, mutex \rangle \] where:
  \begin{itemize}
  \item $tid\in TID$ is the thread ID;
  \item $osl\in OSL$ is the offset-span label;
  \item $bl\in \N$ is the barrier label of the last barrier seen by the thread
    $t$;
  \item $addr\in ADDR$ is the memory address;
  \item $type\in\{R,W\}$ records reads or writes;
  \item $mutex$ is the synchronization state (value of $M$ in $GS$) at the
    time of the access;
  \end{itemize}
\item $\sigma\in\Sigma$ is the data state of the system, as described earlier.
\end{itemize}
\hfill \break

The local state $TP$ is the thread pool that contains a list of 3-tuples,
each one of which is the local state of a thread:
\[ \langle tid, osl, bl \rangle \]

The domain $TP=2^{TID\times OSL\times \N}$ where:

\begin{itemize}
\item $t\in TID$ is the id of the thread;
\item $osl\in OSL$ is an offset-span label;
\item $bl\in\N$ is the label of the barrier the thread has witnessed last.
  We assume that each barrier instruction is of the form $bar(L)$ where
  $L\in\N$ carries the barrier number.
  A thread crossing a barrier sets its $bl$ to the value $L$.
\end{itemize}

\subsection{Helper Functions and Predicates}
\label{sec:helperspredicates}

We define some helper functions to support the operational semantics rules.
They can be operators or functions that receive some arguments in input and
return a certain result or state useful for the rule execution.
The helper functions are the following:

\begin{itemize}
\item $as$: is used as in Ocaml (it allows a name for a whole structure, as
  well as helps us refer to the inner details of the structure).
\item $most(lst)$: we define $most$ as a function that returns the same list
  given in input except the last element (i.e.\ in Python lst[:-1]).
\item $\parallel$: This operator is used to describe that two different
  threads are concurrent.
  In particular, given two offset-span labels $osl_1$ for thread $T_1$ and
  $osl_2$ for thread $T_2$, $osl_1 \parallel osl_2$ (read $osl_1$ and $osl_2$
  are concurrent) means that the threads $T_1$ and $T_2$ may race.
\item $SpawnChildren(\langle ptid, posl, pbl \rangle, \sigma, N)$: Given the
  parent's thread id ($ptid$), offset-span label ($posl$) and barrier label
  ($pbl$), this function creates a pool of $N$ threads --- specifically, the
  local states of these threads $\langle tid, osl, bl \rangle$.
  It initializes the offset-span label $osl$ for each thread created (e.g.\
  at the beginning of a parallel region), by extending $posl$ with pairs
  $[0,N]$ through $[N-1,N]$.
  The $bl$ is set to $pbl$.
  The threads id are somehow uniquely generated.
\item $GetChildJoin(tp)$: returns the single thread-state triple that result
  from fusing all the threads in the thread pool $tp$.
\item $Concurrent(OSL, t_1, t_2)$ is the function that compares the
  offset-span labels as described in Section~\ref{sec:osl}.
\item $AddRW(\langle tid, osl, bl, addr, mat, m, n \rangle)$ adds the access
  into the {\em rw} structure.
  The record says ``an access by $tid$ with offset-span label $osl$ and
  barrier label $bl$ is performed at address $addr$ with memory access type
  $mat$, when the mutex state is $m$.''
\item $Full(bm, osl)$: This predicate keeps the count of the number of threads
  that have reached a $ParEnd(N)$ (or a $Barrier(bid)$) construct.
  In order to count the threads, it uses the structure $bm$ which is indexed
  by the $ParRegID$ represented by the offset-span label $osl$.
  In other words, the predicate $Full$ means that other threads have reached
  the construct and have incremented the counter in the $bm$ structure.
  From a functional language point of view $Full$ would look like:

  \begin{center}
    \lstset{language=[Objective]Caml}
    \begin{lstlisting}
      let Full(bm, osl) =
         let (count, N) = bm[osl]
         in (count == N - 1)
    \end{lstlisting}
  \end{center}

\item $WaitAtBarrier(bid)$: This predicate is used for the example in
  Section~\ref{sec:walkthroughexample} to indicate that a thread already
  encountered a barrier and it is waiting for the other threads in the team.
\item $RaceFail(state, addr, t_1, t_2)$: This helper function is used to
  report the race found on {\em addr}, between thread $t_1$ and thread $t_2$.
\end{itemize}

\subsection{Operational Semantics Rules}
\label{sec:ruleswalkthrough}

Now, we explain the rules in Table~\ref{fig:opsem} one by one.
While each rule models a different behavior, all rules update the system state
incrementing the program counter to point to the next instruction.

\paragraph*{Parallel Region Begin} The \emph{ParallelBegin} rule models the
creation of the team of threads for the encountered parallel region and
initializes the offset-span labels for each thread.
\paragraph*{Parallel Region End} The \emph{ParallelEnd} rule models the end of
the parallel region. It terminates the threads in the team except the master
thread which resumes its execution.
\paragraph*{Implicit Task Begin} The \emph{ImplicitTaskBegin} rule fires when
a thread, after its creation, begins the associated implicit task which
performs the work within the parallel region.
This rule is a helper transition to initialize the thread and its implicit
task state.
\paragraph*{Implicit Task End} The \emph{ImplicitTaskEnd} fires when a thread
exits the implicit barrier and the parallel region is terminating.
It also resets the thread state.
\paragraph*{Load Store} The \emph{LoadStore} rule triggers every time a thread
performs a read or a write operation.
Its task is to store the information about the current memory accesses of a
thread along with other information such as the current locks held by the
task, offset-span label, etc.
The information about a load or a store are kept in a data structure shared
among all threads.
\paragraph*{Acquire Mutex} The rule \emph{AcquireMutex} fires when a thread
encounters a synchronization construct, such as a critical section.
It stores the id ($\mu$ in case of global critical section) of the
synchronization construct into a data structure for the given thread.
All the following memory accesses are stored with the information that
they happened within the given synchronization region.
\paragraph*{Release Mutex} The rule \emph{Release Mutex} instead fires when a
thread encounters the end of a critical section or release a lock.
It removes, from the thread's data structure, the id of the synchronization
construct.
\paragraph*{Barrier} The \emph{Barrier} rules are of extreme importance since
they implement the data race detection algorithm.
The first two rules make sure that all threads in a team reached the barrier
and update the information in the global state.
Once all threads have hit the current barrier the third rule triggers and
perform the race check.
The data race check consists of searching for memory accesses conflicts
between each given pair of concurrent threads.
First, the rule checks if the pair contains two concurrent threads, either
checking if they belong to the same barrier interval or comparing the
offset-span labels.
In the event the threads are concurrent, the rule applies the other checks to
search for data races.
It looks into the loads/stores data structures for memory accesses with the
same address, checks if at least one of them is a write and they do not have
any synchronization regions in common.
In case all these checks are positive the rule triggers a \emph{RaceFail}
event to report the data race.

\input{transitiontable}

\subsection{Operational Semantics Example}
\label{sec:walkthroughexample}

In this section we show an application of the operational semantics in an
OpenMP example.
We show how each rule is triggered according to the operations performed by
the program.
We also provide a transition table to illustrate the system state and how it
changes under the execution of each rule.
The example we use is the OpenMP program shown in
Listing~\ref{code:example01}.
Initially we have only the main thread, the total state of the system is
therefore the following:

\[
init = \langle gs, tp \rangle
\]

with:

\[ gs = \langle bm, m, rw, \sigma \rangle \in GS \]
\[ tp = \langle tid, osl, bl \rangle \in TP \]

where:

\[ gs = \langle \emptyset, \emptyset, \emptyset, \sigma \rangle \]
\[ tp = \langle (0, [0, 1], 0) \rangle \]

\begin{figure}[!ht]
  \centering
  \includegraphics[width=0.37\textwidth]{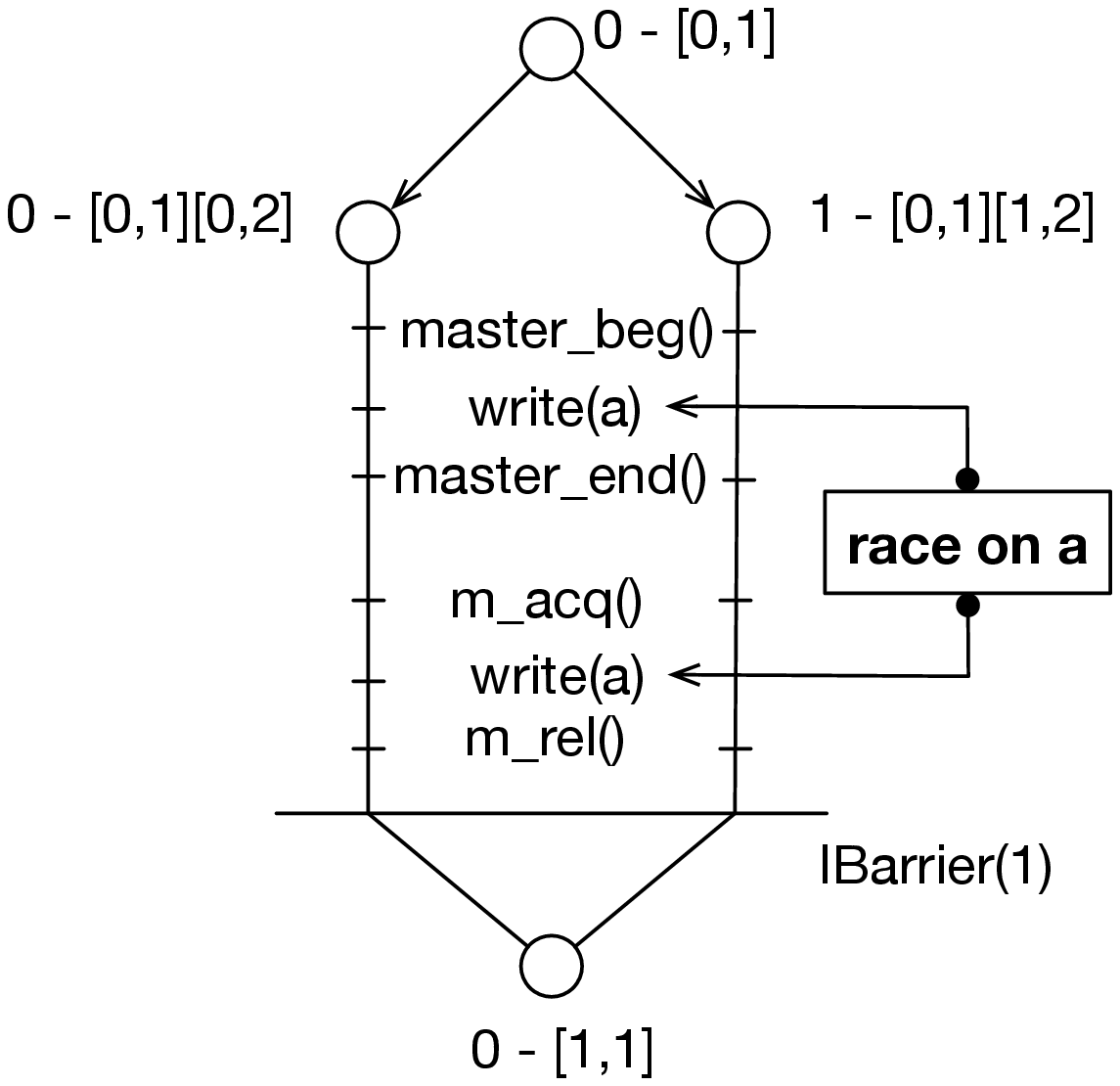}
  \caption{Structure of the OpenMP program in Listing~\ref{code:example01}.}
  \label{fig:example01}
\end{figure}

The Table~\ref{tab:transitions} illustrates the transition table of the system
for the example in Figure~\ref{fig:example01}.
Each thread in the table is represented by its thread id and offset-span
label.

The row 0 of the transition table shows the initial state of the system.
The first fired rule is {\em ParBegin(2)} (Row 1) when the thread 0 hits the
parallel construct.
This rule models the beginning of the parallel region and the creation of the
team of threads.
In the example, the master thread creates one more thread to make a team of
two.
Both threads in the system trigger the {\em ImplicitTaskBegin} rule (Row 2
and 3) to initialize their status (e.g.\ offset-span labels, state, barrier
counts, etc.).
Now the threads start their parallel work.
Thread 0 triggers the {\em LoadStore} rule (Row 4) when it accesses the master
construct and initializes the variable {\em a}.
The rule adds the memory access information inside the {\em rw} data structure
and points to the next instruction.
In the next instruction, thread 0 acquires the mutex which triggers the {\em
  AcquireMutex} rule (Row 5) and updates the thread state with the
synchronization information.
Thread 0 accesses again variable {\em a} and the {\em LoadStore} rule (Row 6)
adds the new memory access to {\em rw} along with the synchronization
information acquired by the previous operation.
The thread 0 releases the mutex triggering the {\em ReleaseMutex} rule (Row 7)
and reaches the implicit barrier at the end of the parallel region.
The triggering of the {\em Barrier} rule (Row 7) keeps thread 0 on waiting for
thread 1 to reach the barrier.

Thread 1 triggers respectively {\em AcquireMutex}, {\em LoadStore}, and {\em
  ReleaseMutex} (Row 9, 10, 11), which add a new synchronized memory access
into the {\em rw} data structure.
Now thread 1 reaches the implicit barrier triggering the {\em Barrier} rule
(Row 12).
The {\em Barrier} rule performs the data race detection which identifies the
data race between the non-synchronized access from thread 0
($\langle 0, [0,1][0,2], [0,1][0,2][0], x, W, \emptyset \rangle$) and the
synchronized access from thread 1
($\langle 1, [0,1][0,2], [0,1][0,2][0], x, W, {\mu} \rangle$).
The two accesses are performed by two different threads in the same memory
location, both happen in the same barrier interval (concurrently according to
offset-span label), at least one of the operations is a write, and one of them
happens outside the critical section $\mu$.
The system reports the race through the {\em RaceFail} helper function.

The execution of the program continues triggering the {\em ImplicitTaskEnd}
rule (Row 13 and 14) by both threads.
Thread 1 terminates immediately, while thread 0 reaches the end of the
parallel region and terminates with the end of the program.

\subsection{Lowering OpenMP constructs}
\label{sec:openmpconstructs}

Our operational semantics models the concurrency structure of an OpenMP
program that uses a subset of the entire OpenMP
specification~\cite{tim_lewis_openmp_nodate}.
We target OpenMP parallel directives and all related constructs except
explicit tasks and target devices that we leave to future works.
Our formalization lowers every OpenMP directive into basic underlying
synchronization structures such as barriers and mutex.
In the following paragraphs, we show how each of these directives can be
simplified and modeled by the operational semantics.

\paragraph{\normalfont \texttt{parallel} \textbf{Construct}} The first five
rules~(\ref{rule:parallelbegin}--\ref{rule:implicittaskend}), in
Figure~\ref{fig:opsem}, model the begin/end of a parallel construct including
the creation and destruction of the implicit task associated to the threads.
The threads within the parallel region trigger the other rules based on the
work they are performing: accessing shared or private
memory~(\ref{rule:loadstore}), acquiring/releasing
mutexes~(\ref{rule:acquiremutex},\ref{rule:releasemutex}), synchronizing to an
implicit/explicit barrier~(\ref{rule:barrier1}--\ref{rule:barrierracecheck}).
The data race detection algorithm performed at the barrier (either implicit or
explicit) catches the potential race(s).
The clauses related to the parallel region constructs do not influence the
data race detection.
For example, in presence of the {\em private} clause or similar, when the
threads access their own private memory, the memory addresses of the locations
are different for each thread, thus no race is reported.

\paragraph{\normalfont \texttt{worksharing} \textbf{Constructs}} The
worksharing constructs such as {\em for}, {\em section}, {\em single}, and
{\em workshare} are also supported by the operational semantics.
These constructs add an implicit barrier at the end, so the race detection
algorithm runs when the thread synchronizes, identifying any potential race
within the barrier interval.
In the presence of a {\em nowait} clause, the operational semantics models the
specific constructs as an extension of the parallel work until the next
barriers.
Let us take the example in Listing~\ref{code:example03}.
The snippet of code shows two consecutive parallel for-loops with the {\em
  nowait} clause.
The clause removes the implicit barrier at the end of the first parallel loop,
introducing a data dependency between the write on {\em a[i]} in the first
loop and the read on {\em a[i]} and {\em a[i-1]} in the second loop.
Consequently, all memory accesses performed by the threads in both loops
happen in the same barrier interval.
Only at the end of the second loop, when the threads encounter the implicit
barrier, the state machine triggers the data race detection analysis
(Rule~\ref{rule:barrierracecheck}).
In detail, the state machine stores information about the memory locations
accessed by the threads in both loops.
Because of the data dependency between the loops, the race check identifies
two common non-synchronized memory accesses, in the {\em rw} data structure,
from two different threads.
Since one of the accesses is a write, the operational semantics reports the
data race.

\begin{figure}[H]
  \input{example03}
\end{figure}

\paragraph{\normalfont \texttt{master} and \texttt{synchronization}
  \textbf{Constructs}}

The only synchronization constructs not supported by the operational semantics
are those related to tasking: {\em taskwait} and {\em taskgroup} which, as
said previously, will be modeled in future work.
When a thread encounters a synchronization directive, a rule logs the
synchronization information for the current thread.
Every memory access executed by the thread within a synchronization construct
is collected in the {\em rw} data structure, with the information that the
memory access are protected by a synchronization primitive.
The data race detection, as shown in rule~\ref{rule:barrierracecheck}, uses
this information to identify a potential non-synchronized access and report
the race.


%% file: example04.tex
  \lstset{language=C++,
    escapeinside=||,
    basicstyle=\ttfamily,
    showstringspaces=false,
    keywordstyle=\color{blue}\ttfamily,
    stringstyle=\color{black}\ttfamily,
    commentstyle=\color{gray}\ttfamily,
    otherkeywords={omp, parallel, barrier, critical, num_threads, shared},
    breaklines=true,
    basicstyle=\small
  }
\begin{lstlisting}[language=C++, caption=OpenMP program with nested parallel regions., label=code:example04, belowskip=-0.8\baselineskip]
#pragma omp parallel shared(x, y) num_threads(2)
  {
    if(|omp|_get_thread_num() % 2 == 0) {
    // Left-branch of the graph
#pragma omp parallel num_threads(2)
      {
#pragma omp critical
        {
          x = 1;
        }
#pragma omp barrier
        y = x;
      }
#pragma omp parallel num_threads(2)
      {
#pragma omp critical(M1)
        {
          printf("Y: %d\n", y);
        }
      }
    } else {
      //Right-branch of the graph
#pragma omp parallel num_threads(2)
      {
#pragma omp critical(M1)
        {
          y = y + 1;
        }
#pragma omp barrier
#pragma omp for
        for(int i = 0; i < 10; i++) {
#pragma omp critical
          {
            x = x + 1;
          }
        }
      }
    }
  }
\end{lstlisting}


%% file: rules.tex
\begin{figure*}
  \centering
  \caption{OpenMP Concurrency Operational Semantics}
  \label{fig:opsem}
  \small
  \begin{framed}
    \textbf{Operational Semantics State}
    \begin{align}
      gs\; {\rm as}\; (bm, m, rw, \sigma) \in GS \\
      te\; {\rm as}\; (tid, osl, bl) \in TP
    \end{align}
    \textbf{Operational Semantics Rules}
    \begin{align}
      {\bf ParallelBegin(N)} & \inference[]{
                               at(tid, \sigma, ParBegin(N)) \wedge \\
                               tp' = (tp - \{te\} \cup SpawnChildren(\langle tid, osl, bl \rangle, \sigma, N)) \wedge \\
      bm' = bm \cup \{\langle osl,(0,N)\rangle\} \wedge
      \sigma' = nxt(\sigma, tid)}
      {\langle gs, tp \rangle \longrightarrow
      \langle gs'\; {\rm as}\; \langle bm',m,rw,\sigma'\rangle, tp' \rangle} \label{rule:parallelbegin} \\
      \nonumber \\
      {\bf ParallelEnd(N)} & \inference[]{
                             tp' \subseteq tp \wedge
                             at(tid, \sigma, ParEnd(N)) \wedge \\
                             \sigma' = nxt(\sigma, tid) \wedge
                             tp'' = tp - tp' \cup GetChildJoin(tp')}
                             {\langle gs, tp \rangle \longrightarrow
                             \langle gs'\; {\rm as}\; \langle bm,m,rw,\sigma'\rangle, tp'' \rangle} \label{rule:parallelend} \\
      \nonumber \\
      {\bf ImplicitTaskBegin()} & \inference[]{
                                  at(tid, \sigma, ImplicitTaskBegin()) \wedge
                                  \sigma' = nxt(\sigma, tid)}
                                  {\langle gs, tp \rangle \longrightarrow
                                  \langle gs'\; {\rm as}\; \langle
                                  bm,m,rw,\sigma'\rangle, tp \rangle} \label{rule:implicittaskbegin} \\
      \nonumber \\
      {\bf ImplicitTaskEnd()} & \inference[]{
                                at(tid, \sigma, ImplicitTaskEnd()) \wedge
                                \sigma' = nxt(\sigma, tid)}
                                {\langle gs, tp \rangle \longrightarrow
                                \langle gs'\; {\rm as}\; \langle
                                bm,m,rw,\sigma'\rangle, tp \rangle} \label{rule:implicittaskend} \\
      \nonumber \\
      {\bf LoadStore()} & \inference[]{
                          at(tid, \sigma, LoadStore(addr, mat)) \wedge \\
      rw' = ADDR - RW(tid, osl, bl, addr, mat, mutex) \wedge
      \sigma' = nxt(\sigma, tid)}
      {\langle gs, tp \rangle \longrightarrow \langle gs'\; {\rm as}\; \langle
      bm,m,rw',\sigma' \rangle, tp \rangle} \label{rule:loadstore} \\
      \nonumber \\
      {\bf AcquireMutex(name)} & \inference[]{
                                 at(tid, \sigma, AcquireMutex(name)) \wedge
                                 m[name] = \emptyset \wedge \\
      m' = m[name \rightarrow tid] \wedge
      \sigma' = nxt(\sigma, tid)}
      {\langle gs, tp \rangle \longrightarrow \langle gs'\; {\rm as}\; \langle
      bm,m',rw,\sigma' \rangle, tp \rangle} \label{rule:acquiremutex} \\
      \nonumber \\
      {\bf ReleaseMutex(name)} & \inference[]{
                                 at(tid, \sigma, ReleaseMutex(name)) \wedge
                                 m[name] = tid \wedge \\
      m' = m[name \rightarrow \emptyset] \wedge
      \sigma' = nxt(\sigma, tid)}
      {\langle gs, tp \rangle \longrightarrow \langle gs'\; {\rm as}\; \langle
      bm,m',rw,\sigma' \rangle, tp \rangle} \label{rule:releasemutex} \\
      \nonumber \\
      {\bf Barrier(bid)} & \inference[]{
                           at(tid, \sigma, Barrier(bid)) \wedge
                           Full(bm, most(osl)) \wedge \\
      bm' = bm - \{\langle osl,* \rangle\} \wedge
      \sigma' = nxt(\sigma, tid)}
      {\langle gs, tp \rangle \longrightarrow
      \langle gs'\; {\rm as}\; \langle bm',m,rw,\sigma'\rangle, tp \rangle} \label{rule:barrier1} \\
      \nonumber \\
      {\bf Barrier(bid)} & \inference[]{
                           at(tid, \sigma, Barrier(bid)) \wedge
                           \neg Full(bm, most(osl)) \wedge \\
                           bm[most(osl)]\; as\; (count, N) \wedge \\
      te' as (tid, osl, bid) \wedge
      tp' = tp - te \cup \{ te' \} \wedge \\
      bm' = bm \cup \{\langle osl,(count+1,N) \rangle\} \wedge
      \sigma' = nxt(\sigma, tid)}
      {\langle gs, tp \rangle \longrightarrow \langle gs'\; {\rm as}\; \langle
      bm',m,rw,\sigma'\rangle, tp' \rangle} \label{rule:barrier2} \\
      \nonumber \\
      {\bf Barrier(bid)} & \inference[]{
                           te_1\; as\; (tid_1, osl1, bl1) \in tp \wedge
                           te_2\; as\; (tid_2, osl2, bl2) \in tp \wedge
                           (tid_1 \neq tid_2) \wedge \\
      Concurrent(osl, tid_1, tid_2) \wedge
      i \in rw[tid_1] \wedge
      j \in rw[tid_2] \wedge \\
      (rw[tid_1][i].addr == rw[tid_2][j].addr) \wedge \\
      (rw[tid_1][i].mat == W) \wedge (rw[tid_2][j].mat == W) \wedge \\
      (rw[tid_1][i].mutex \cap rw[tid_2][j].mutex = \emptyset) \wedge \\
      (rw[tid_1][i].bl == rw[tid_2][j].bl) \wedge (rw[tid_1][i].bl \parallel rw[tid_2][j].bl)}
      {\langle gs, tp \rangle \rightarrow RaceFail(\sigma, addr, tid_1,
      tid_2)} \label{rule:barrierracecheck}
    \end{align}
  \end{framed}
\end{figure*}


%% file: transitiontable.tex
\begin{table*}[ht]
  \centering
  \caption{State machine transitions for the example in Listing~\ref{code:example01}.}
  \label{tab:transitions}
  \resizebox{1\textwidth}{!}{
    \begin{tabular}{| c | c | c | c | c | c | c | c |}
      \hline
      \textbf{\#} & \textbf{tid - osl} & \textbf{rule} & \textbf{bm} & \textbf{} & \textbf{rw} & \textbf{tp} & \textbf{Next State} \\
      \hline
      \hline
      $0$ & $Init$ & --- & $\emptyset$ & $\emptyset$ & $\emptyset$ & $\langle 0, [0,1], 0 \rangle$ & $ParBegin(2)$ \\
      \hline
      \hline
      $1$ & $0 - [0,1]$ & $ParBegin(2)$ & ${[0,1] = (0, 2)}$ & $\emptyset$ & $\emptyset$ & \specialcell{$\langle 0, [0,1][0,2], 0 \rangle$\\$\langle 1, [0,1][1,2], 0 \rangle$} & \\
      \hline
      \hline
      $2$ & $0 - [0,1][0,2]$ & $ImplicitTaskBegin()$ & \specialcell{$[0,1] = (0, 2)$\\$[0,1][0,2] = (0, 2)$} & $\emptyset$ & $\emptyset$ & \specialcell{$\langle 0, [0,1][0,2], 0 \rangle$\\$\langle 1, [0,1][1,2], 0 \rangle$} & \\
      \hline
      \hline
      $3$ & $1 - [0,1][1,2]$ & $ImplicitTaskBegin()$ & \specialcell{$[0,1] = (0, 2)$\\$[0,1][0,2] = (0, 2)$\\$[0,1][1,2] = (0, 2)$} & $\emptyset$ & $\emptyset$ & \specialcell{$\langle 1, [0,1][1,2], 0 \rangle$} & \\
      \hline
      \hline
      $4$ & $0 - [0,1][0,2]$ & $LoadStore(x, W)$ & \specialcell{$[0,1] = (0, 2)$\\$[0,1][0,2] = (0, 2)$\\$[0,1][1,2] = (0, 2)$} & $\emptyset$ & $\langle 0, [0,1][0,2], [0,1][0,2][0], x, W, \emptyset \rangle$ & \specialcell{$\langle 0, [0,1][0,2], 0 \rangle$\\$\langle 1, [0,1][1,2], 0 \rangle$} & $AcquireMutex()$ \\
      \hline
      $5$ & --- & $AcquireMutex()$ & \specialcell{$[0,1] = (0, 2)$\\$[0,1][0,2] = (0, 2)$\\$[0,1][1,2] = (0, 2)$} & $\emptyset$ & $\langle 0, [0,1][0,2], [0,1][0,2][0], x, W, \emptyset \rangle$ & \specialcell{$\langle 0, [0,1][0,2], 0 \rangle$\\$\langle 1, [0,1][1,2], 0 \rangle$} & $LoadStore(x, W)$ \\
      \hline
      $6$ & --- & $LoadStore(x, W)$ & \specialcell{$[0,1] = (0, 2)$\\$[0,1][0,2] = (0, 2)$\\$[0,1][1,2] = (0, 2)$} & $\emptyset$ & \specialcell{$\langle 0, [0,1][0,2], [0,1][0,2][0], x, W, \emptyset \rangle$\\$\langle 0, [0,1][0,2], [0,1][0,2][0], x, W, {\mu} \rangle$} & \specialcell{$\langle 0, [0,1][0,2], 0 \rangle$\\$\langle 1, [0,1][1,2], 0 \rangle$} & $ReleaseMutex()$ \\
      \hline
      $7$ & --- & $ReleaseMutex()$ & \specialcell{$[0,1] = (0, 2)$\\$[0,1][0,2] = (0, 2)$\\$[0,1][1,2] = (0, 2)$} & $\emptyset$ & \specialcell{$\langle 0, [0,1][0,2], [0,1][0,2][0], x, W, \emptyset \rangle$\\$\langle 0, [0,1][0,2], [0,1][0,2][0], x, W, {\mu} \rangle$} & \specialcell{$\langle 0, [0,1][0,2], 0 \rangle$\\$\langle 1, [0,1][1,2], 0 \rangle$} & $Barrier(1)$ \\
      \hline
      $8$ & --- & $Barrier(1)$ & \specialcell{$[0,1] = (0, 2)$\\$[0,1][0,2] = (1, 2)$\\$[0,1][1,2] = (2, 2)$} & $\emptyset$ & \specialcell{$\langle 0, [0,1][0,2], [0,1][0,2][0], x, W, \emptyset \rangle$\\$\langle 0, [0,1][0,2], [0,1][0,2][0], x, W, {\mu} \rangle$} & \specialcell{$\langle 0, [0,1][0,2], 0 \rangle$\\$\langle 1, [0,1][1,2], 0 \rangle$} & \specialcell{$WaitAtBarrier(1)$\\$ImplicitTaskEnd()$} \\
      \hline
      \hline
      $9$ & $1 - [0,1][0,2][0,2]$ & $AcquireMutex()$ & \specialcell{$[0,1] = (0, 2)$\\$[0,1][0,2] = (0, 2)$\\$[0,1][1,2] = (0, 2)$} & $\emptyset$ & \specialcell{$\langle 0, [0,1][0,2], [0,1][0,2][0], x, W, \emptyset \rangle$\\$\langle 0, [0,1][0,2], [0,1][0,2][0], x, W, {\mu} \rangle$} &  \specialcell{$\langle 1, [0,1][1,2], 0 \rangle$} & $LoadStore(x, W)$ \\
      \hline
      $10$ & --- & $LoadStore(x, W)$ & \specialcell{$[0,1] = (0, 2)$\\$[0,1][0,2] = (0, 2)$\\$[0,1][1,2] = (0, 2)$} & $\emptyset$ & \specialcell{$\langle 0, [0,1][0,2], [0,1][0,2][0], x, W, \emptyset \rangle$\\$\langle 0, [0,1][0,2], [0,1][0,2][0], x, W, {\mu} \rangle$\\$\langle 1, [0,1][0,2], [0,1][0,2][0], x, W, {\mu} \rangle$} &  \specialcell{$\langle 1, [0,1][1,2], 0 \rangle$} & $ReleaseMutex()$ \\
      \hline
      $11$ & --- & $ReleaseMutex()$ & \specialcell{$[0,1] = (0, 2)$\\$[0,1][0,2] = (0, 2)$\\$[0,1][1,2] = (0, 2)$} & $\emptyset$ & \specialcell{$\langle 0, [0,1][0,2], [0,1][0,2][0], x, W, \emptyset \rangle$\\$\langle 0, [0,1][0,2], [0,1][0,2][0], x, W, {\mu} \rangle$\\$\langle 1, [0,1][0,2], [0,1][0,2][0], x, W, {\mu} \rangle$} &  \specialcell{$\langle 1, [0,1][1,2], 0 \rangle$} & $Barrier(1)$ \\
      \hline
      $12$ & --- & $Barrier(1)$ & \specialcell{$[0,1] = (0, 2)$\\$[0,1][1,2] = (1, 2)$\\$[0,1][0,2] = (2, 2)$} & $\emptyset$ & \specialcell{$\langle 0, [0,1][0,2], [0,1][0,2][0], x, W, \emptyset \rangle$\\$\langle 0, [0,1][0,2], [0,1][0,2][0], x, W, {\mu} \rangle$\\$\langle 1, [0,1][0,2], [0,1][0,2][0], x, W, {\mu} \rangle$} & \specialcell{$\langle 1, [0,1][1,2], 0 \rangle$} & $RaceFail(\sigma, x, 0, 1)$ \\
      \hline
      $13$ & --- & $ImplicitTaskEnd()$ & $\emptyset$ & $\emptyset$ & $\emptyset$ & \specialcell{$\langle 1, [0,1][1,2], 0 \rangle$} & \\
      \hline
      \hline
      $14$ & $0$ & $ImplicitTaskEnd()$ & $\emptyset$ & $\emptyset$ & $\emptyset$ & \specialcell{$\langle 0, [0,1][0,2], 0 \rangle$} & $ParEnd(2)$ \\
      \hline
      $15$ & --- & $ParEnd(2)$ & $\emptyset$ & $\emptyset$ & $\emptyset$ & $\langle 0, [1,1], 0 \rangle$ & \\
      \hline
    \end{tabular}}
\end{table*}


%% file: example03.tex
\lstset{language=C++,
  escapeinside=||,
  basicstyle=\ttfamily,
  showstringspaces=false,
  keywordstyle=\color{blue}\ttfamily,
  stringstyle=\color{red}\ttfamily,
  otherkeywords={omp, parallel, barrier, critical, num_threads, shared, master},
  breaklines=true,
  basicstyle=\small
}
\begin{lstlisting}[language=C++, caption=Data race on array {\em a} because of
{\em nowait} clause and data dependency between two for loops.,
label=code:example03, belowskip=-0.5\baselineskip, aboveskip=-0.5\baselineskip]
#pragma omp parallel
  {
#pragma omp for nowait
    for (i = 0; i < N; i++) {
      a[i] = 3.0 * i * (i + 1);;
    }
#pragma omp for
    for (i = 1; i < N; i++) {
      b[i] = a[i] - a[i - 1];
    }
  }
\end{lstlisting}


%% file: implementation.tex
\section{Implementation}
\label{sec:implementation}

The operational semantics is a mathematical model and must clearly be adapted
to real-world implementation settings.
We have implemented a preliminary version of such a tool called \sword.
The main idea behind this tool is to log all OpenMP events and memory accesses
into a file (one such file is created per thread).
When the program execution terminates, an offline data race detection
algorithm analyzes the log files to identify potential data races.
The main advantages of this approach are: (1)~dramatically reduced memory
overheads compared to other tools (including \archer), and (2)~parallelizable
offline analysis.

More specifically, \sword includes a compiler instrumentation pass for the
source program and two checking phases.
The compiler instrumentation inserts in the program, for each load and store,
a call to a \sword runtime routine that implements the event collection
algorithm.
Phase one consists of logging into files every memory access and
synchronization operation that each thread executes at runtime.
The \sword runtime intercepts parallel regions begin/end, synchronization
operations (e.g.\ critical sections, barriers, etc.), and other OpenMP events
through the OMPT interface.
This implementation benefits from our operational semantics directly including
events that match OMPT events.

During the execution of the program, the \sword runtime uses a buffer for each
thread to collect the data regarding memory accesses and OpenMP events.
When the buffer is full, \sword compresses it, dumps it in a log file, and
makes it available to collect new data.
The use of data compression in this manner helps reduce memory overheads.
Once the program finishes its execution, the log folder contains a log-file
per thread.

The second phase consists of the offline analysis of the logs to identify the
data races that manifested during the program execution.
%
%
The algorithm identifies the pairs of concurrent threads using the offset-span
label mechanism described in Section~\ref{sec:osl}.
The data race detection algorithm identifies memory conflicts between two
concurrent threads.
The algorithm obtains the information about the thread's memory accesses and
synchronization operations from the logs, and looks for data races.
Since the analysis requires only to read from the log files, the offline
algorithm can be parallelized across multiple cores and a cluster of nodes to
speedup the process.


%% file: conclusions.tex
\section{Conclusions}
\label{sec:conclusions}

In this paper, we have presented an operational semantics to model the
concurrency structure of OpenMP and enabling data race detection for
structured parallelism.
The operational semantics rules are straightforward and can serve as a
valuable reference to everyday programmers.
Also, the example~\ref{sec:walkthroughexample} shows how our approach can
identify data races even in corner cases where other techniques (e.g., those
purely based on the happens-before tracking) can fail.
In summary, our work provides a formalization to help researchers and tool
developers to better understand OpenMP concurrency, and help them reliably and
systematically build more precise data race checkers that reduce memory
overheads.

As already described, we are working on a possible implementation of the
operational semantics to support a new data race checker called \sword.
Details of the engineering of \sword will be presented in future work.

To the best of our knowledge, our contribution is the first simple operational
semantics to model the concurrency structure of OpenMP at a level that
tool-builders care about.
Our semantics is not yet suitable for those interested in issues such as
(1)~OpenMP's weak memory consistency model, (2)~OpenMP's GPU offload features,
and (3)~OpenMP's tasking constructs.
However, our semantics offers a very appealing starting point for such
extensions.

The operational semantics rules mesh with the OMPT events providing a powerful
as well as {\em standardized} instrumentation approach to represent the
concurrency structure of an OpenMP program and enable targeted data race
detection.
We believe that with this formalization and the ongoing work we can build
precise and accurate data race checkers that exploit the structured
parallelism of parallel programming models such as OpenMP and its future
incarnations.
